\documentclass[modern]{aastex63}

\usepackage{amsmath}
\usepackage{amssymb}
\usepackage{rotating}

\newcommand\tE{t_{\rm E}}
\newcommand\tEgeo{t_{\rm E}^{\rm geo}}

\newcommand\piEN{\pi_{\textrm{E},N}}
\newcommand\piEE{\pi_{\textrm{E},E}}

\newcommand\tpar{t_{\rm 0,par}}
\newcommand\muvec{\boldsymbol{\mu}}
\newcommand\mugeo{\boldsymbol{\mu}^{\rm geo}}
\newcommand\muhelio{\boldsymbol{\mu}^{\rm bary}}
\newcommand\muhelioS{\boldsymbol{\mu}^{\rm bary}_{\rm source}}

\newcommand\muhelioL{\boldsymbol{\mu}^{\rm bary}_{\rm lens}}
\newcommand\ve{\boldsymbol{v_{\oplus,\perp}}}
\newcommand\pirel{\pi_{\rm rel}}
\newcommand\thetaE{\theta_{\rm E}}
\newcommand\piE{\pi_{\rm E}}
\newcommand\piEvec{\boldsymbol{\pi_{\rm E}}}
\newcommand\au{\mathrm{au}}
\newcommand\model{\boldsymbol{\omega}}
\newcommand\data{\boldsymbol{d}}

\defcitealias{wyrzykowski2020b}{WM20}
\defcitealias{wyrzykowski2016}{W16}

\received{}
\revised{}
\accepted{}

\shorttitle{Masses of dark stellar remnants}
\shortauthors{P. Mr\'oz \& \L{}. Wyrzykowski}

\begin{document}

\title{Measuring the mass function of isolated stellar remnants with gravitational microlensing. I. Revisiting the OGLE-III dark lens candidates}

\correspondingauthor{Przemek Mr\'oz}
\email{pmroz@astro.caltech.edu}

\author[0000-0001-7016-1692]{Przemek Mr\'oz}
\affil{Division of Physics, Mathematics, and Astronomy, California Institute of Technology, Pasadena, CA 91125, USA}
\affil{Astronomical Observatory, University of Warsaw, Al. Ujazdowskie 4, 00-478 Warszawa, Poland}
\author[0000-0002-9658-6151]{\L{}ukasz Wyrzykowski}
\affil{Astronomical Observatory, University of Warsaw, Al. Ujazdowskie 4, 00-478 Warszawa, Poland}

\begin{abstract}
Gravitational microlensing may detect dark stellar remnants -- black holes or neutron stars -- even if they are isolated. However, it is challenging to estimate masses of isolated dark stellar remnants using solely photometric data for microlensing events. A recent analysis of OGLE-III long-timescale microlensing events exhibiting the annual parallax effects claimed that a number of bright events were due to ``mass-gap'' objects (with masses intermediate between those of neutron stars and black holes).
Here, we present a detailed description of the updated and corrected method that can be used to estimate masses of dark stellar remnants detected in microlensing events given the light curve data and the proper motion of the source. We use this updated method, in combination with new proper motions from \textit{Gaia}~EDR3, to revise masses of dark remnant candidates previously found in the OGLE-III data. We demonstrate that masses of ``mass-gap'' and black hole events identified in the previous work are overestimated and, hence, these objects are most likely main-sequence stars, white dwarfs, or neutron stars.
\end{abstract}


\section{Introduction} \label{sec:intro}

Hundreds of millions of isolated stellar remnants -- black holes and neutron stars -- are expected to reside in the Milky Way \citep[e.g.,][]{shapiro1983}. Detecting isolated black holes is especially challenging since they emit little or no electromagnetic radiation\footnote{Isolated black holes may emit X-ray radiation powered by accretion from the interstellar medium \citep[e.g.,][]{shvartsman1971,agol2002}, however, searches for such objects came out empty \citep[e.g.,][]{revnivtsev2002,chisholm2003,maeda2005,nucita2006}.}. One method that is particularly suited for finding isolated stellar remnants is gravitational microlensing -- microlensing occurs when gravity of a lensing object, irrespective of its brightness, bends and magnifies the light of a distant stellar source. An observer on Earth may detect a transient brightening of the source.

Modern surveys for gravitational microlensing events (Optical Gravitational Lensing Experiment, OGLE, \citealt{udalski2015}; Korea Microlensing Telescope Network, KMTNet, \citealt{kim2016}; Microlensing Observations in Astrophysics, MOA, \citealt{sumi2013}) are detecting 2000--3000 events toward the Galactic center every year and \citet{gould2000} estimated that about 4\% of them should be due to remnants (black holes---1\%, and neutron stars---3\%). However, only several black hole candidates were detected using microlensing \citep{bennett2002, mao2002, dong2007, wyrzykowski2016, wyrzykowski2020b} with no unambiguous discovery so far.

The reason behind this discrepancy is the difficulty in measuring masses of lenses in single-lens microlensing events using solely photometric observations and so distinguishing black-hole events from those due to ordinary stars. The only physical parameter that can be measured for every lensing event is its Einstein timescale $\tE$ which depends on the lens mass $M,$ distance $D_l$, and the relative lens-source proper motion $\mu$.
The lens mass can be measured if two additional quantities, angular Einstein radius $\thetaE$ and microlensing parallax $\piE$ are known\footnote{Another possible way of determining the lens mass involves detecting its light in high-resolution images taken after the event, once the lens and source separate in the sky, but this method is not applicable for dark stellar remnants.}:
\begin{equation}
M = \frac{\thetaE}{\kappa\piE},
\label{eq:1}
\end{equation}
where $\kappa=8.144$\,mas\,$M_{\odot}^{-1}$ is a numerical constant \citep{gould2000_nat}. 

The microlensing parallax can be measured from the light curve of the event if it is long enough that the motion of Earth around the Sun induces subtle deviations in the symmetric magnification pattern (the so-called annual parallax effect, \citealt{gould1992}). The amplitude of the parallax effect $\piE=\pirel/\thetaE\propto M^{-1/2}$ (where $\pirel$ is the relative lens-source parallax) is expected to decrease as the mass of the lens gets higher \citep[e.g.,][]{lam2020,karolinski2020}.

Measuring $\thetaE$ using only photometric observations \citep[through finite-source effects;][]{nemi1994,gould1994,witt1994} is rarely possible for single-lens events because it requires special geometry in which the lens is passing almost directly in front of the disk of the source star. Moreover, the occurrence of the effect is biased toward low-mass lenses \citep{gould2013,kim2020}. The angular Einstein radius can be also measured using either interferometric \citep[when two images generated by lensing can be resolved;][]{delplancke2001,cassan2016,dong2019} or astrometric \citep[when the motion of the centroid of the source image is detected;][]{hog1995,miyamoto1995,walker1995,lu2016,rybicki2018} observations. Such observations are currently possible only for a small fraction of events.

\citet{wyrzykowski2016} (hereafter \citetalias{wyrzykowski2016}) carried out a search for microlensing events exhibiting the parallax effect using data from the OGLE-III survey \citep{udalski2003} with the main aim of identifying stellar remnant lenses. They selected a sample of long-timescale events (because $\tE = \thetaE/\mu \propto \sqrt{M}$) that showed the pronounced light curve modulation due to the annual parallax effect and little amount of blended light (which may come from unrelated ambient stars or the lens itself). They found 13 microlensing events which are consistent with having a white dwarf, neutron star, or black hole lens. \citetalias{wyrzykowski2016} were unable to determine precise masses of the lenses because their angular Einstein radius could not be measured. They used the fact that Equation~(\ref{eq:1}) can be rewritten as:
\begin{equation}
M = \frac{\thetaE}{\kappa\piE}=\frac{\tE\mu}{\kappa\piE}.
\label{eq:2}
\end{equation}
Two quantities in this expression can be directly measured from the light curve ($\tE$ and $\piE$), whereas $\mu$ can be constrained from the Milky Way model under the assumption that the velocity distribution of stellar remnants is identical to that of other stars in the Milky Way.

In a follow-up study, \citet{wyrzykowski2020b} (hereafter \citetalias{wyrzykowski2020b}) used additional information from \textit{Gaia}~DR2 \citep{gaia2016,gaia2018} -- distances based on \citet{bailerjones2018} and proper motions of sources -- to further constrain the possible mass distributions of the detected remnant candidates. They found that the derived distribution of masses of dark lenses is consistent with a continuous distribution of stellar-remnant masses (disfavoring the so-called ``mass gap'' between neutron stars and black holes). The estimated mass of the most massive black hole candidate, OGLE3-ULENS-PAR-02, was $11.9^{+4.9}_{-5.2}\,M_{\odot}$.

To estimate the lens mass and distance distribution, \citetalias{wyrzykowski2016} and \citetalias{wyrzykowski2020b} calculated the posterior distribution of five microlensing parameters describing the shape of the light curve. They assigned each MCMC link a random lens-source proper motion and calculated the corresponding lens mass and distance. Subsequently, they assigned each MCMC link a weight equal to:
\begin{equation}
\frac{d^4\Gamma}{d\tE d\mu d^2\piEvec} = \frac{4}{\au}\nu f(\muvec) [g(M)M] \frac{D_l^4\mu^4 \tE}{\piE},
\label{eq:jac}
\end{equation}
where $\nu$ is the stellar density distribution, $f(\muvec)$ is the two-dimensional probability function for a given lens-source relative proper motion $\muvec$, and $g(M)$ is the mass function of lenses. This expression was first derived by \citet{batista2011} and represents the Jacobian of the transformation from the ``physical'' coordinate system ($M, D_l, \muvec$) to the ``natural microlensing parameters'' ($\tE,\mu,\piEvec$) that are measured from the light curve.

Upon a closer examination of the code of \citetalias{wyrzykowski2020b}, we noticed several issues: 1) they assumed that the lens is located in the Milky Way disk and neglected Galactic bulge lenses, 2) they sampled the relative lens-source proper motions from a non-uniform distribution that was skewed toward higher $\mu$ (that is, higher masses), 3) proper motions in heliocentric and geocentric coordinate systems were mixed up, 4) distances to source stars located toward the Galactic bulge taken from \citet{bailerjones2018} were underestimated. These issues combined led to the overestimation of the lens masses.

The main aim of our paper is to provide a detailed description of the updated algorithm that can be used to estimate masses of dark stellar remnants detected in microlensing events and to revise the masses of OGLE-III dark remnant candidates. We also use the updated \textit{Gaia} astrometric measurements from the \textit{Gaia} Early Data Release 3 (EDR3) \citep{gaia_edr3}. We show that OGLE3-ULENS-PAR-02 is most likely a $0.9\,M_{\odot}$ star in the Galactic bulge. We also demonstrate that masses of ``mass-gap'' events claimed by \citetalias{wyrzykowski2020b} are overestimated. Hence, these objects are most likely main-sequence stars, white dwarfs, or neutron stars. 

\section{Methods}

\subsection{Model}

Our model has eight parameters $\model = (t_0,u_0,\tE,\piEN,\piEE,\mu,\mu_{s,N},\mu_{s,E})$. The first five of them describe the shape of the light curve: $t_0$ -- time of the minimal separation between the lens and the source, $u_0$ -- lens--source angular separation at $t_0$ (expressed in $\thetaE$ units), $\tE$ -- Einstein timescale, $\piEN,\piEE$ -- North and East components of the microlens parallax vector. Following \cite{gould2004}, these quantities are measured in the geocentric frame that is moving with a velocity equal to that of the Earth at fiducial time $t_{0,\rm par}$ (which is usually chosen close to $t_0$). The sixth parameter, $\mu$, is the relative lens-source proper motion measured in the same geocentric frame (note that $\mu = |\muvec|$ is a scalar quantity). The last two parameters, $\mu_{s,N}$ and $\mu_{s,E}$, are the North and East components of the source proper motion vector (relative to the solar system barycenter).

From Bayes' Theorem, the posterior distribution $p(\model|\data)$ of model parameters $\model$ given the light curve data $\data$ is:
\begin{equation}
p(\model|\data) = \frac{p(\data|\model)p(\model)}{p(\data)},
\label{eq:posterior}
\end{equation}
where $p(\data|\model)$ is the likelihood function, $p(\model)$ is the prior distribution, and $p(\data)=\int p(\data|\model)p(\model)d\model$. We use $\chi^2$ as our likelihood function:
\begin{equation}
p(\data|\model) = \exp(-\chi^2/2),
\end{equation}
where
\begin{equation}
\chi^2 = \sum_i \left(\frac{F_i-F_{i,\rm model}}{\sigma_i}\right)^2
\end{equation}
and $(F_i,\sigma_i)$ are the $i$th data point and $F_{i,\rm model}$ is the brightness of the event predicted by the microlensing model (as a function of $t_0$, $u_0$, $\tE$, $\piEN$, and $\piEE$). Our prior is the product of the ``Galactic prior'' derived by \citet{batista2011} (Equation~(\ref{eq:jac})) and priors on the source proper motion (taken from \textit{Gaia}):
\begin{equation}
p(\model) = f_{\rm Gal}(\model) f(\mu_{s,N},\mu_{s,E}),
\label{eq:prior}
\end{equation}
where:
\begin{equation}
f_{\rm Gal}(\model) = \frac{4}{\au}(\nu_d f_d(\muvec) + \nu_b f_b(\muvec))[g(M)M]\frac{D_l^4\mu^4 \tE}{\piE}.
\label{eq:galprior}
\end{equation}
Here, $\nu_d$ and $\nu_b$, and $f_d(\muvec)$ and $f_b(\muvec)$ are densities and proper motion distributions of disk and bulge lenses, respectively. Following \citetalias{wyrzykowski2016}, we assume $g(M) \propto M^{-1.75}$. We also keep the source distance $D_s=8$\,kpc fixed. However, $D_s$ can be also regarded as an additional model variable, provided that a prior $f(D_s)$ is also included in Equation~(\ref{eq:prior}), as discussed in Section~\ref{sec:dist}.

\subsection{Posterior distribution of the relative lens-source proper motion}

Let us consider a simpler, one-parameter light curve model that depends only on the relative lens-source proper motion $\mu$. We keep the microlensing parameters $t_0,u_0,\tE,\piEN,\piEE$ fixed at their best-fitting values and we also treat the source proper motion as known and fixed. From Equation~(\ref{eq:posterior}), the posterior distribution of $\mu$ depends only on the Galactic prior:
\begin{equation}
f(\mu | \data) \propto f_{\rm Gal} (\mu),
\end{equation}
which is unsurprising because the light curve shape does not explicitly depend on the value of $\mu$\footnote{The value of $\mu$ can be constrained from the light curve provided that finite-source effects are present. In the case of single-lens events, finite-source effects can be detected only in high-magnification events. The sample analyzed here does not contain such events.}. 

To illustrate the example posterior distributions of $\mu$ and the evaluation of the Galactic prior, we use the best-fitting solution of OGLE3-ULENS-PAR-02 taken from Table~3 of \citetalias{wyrzykowski2016}. This solution has the following three physical parameters (calculated in the geocentric frame at $\tpar = 2454018.6$, that is October 10.1, 2006): $\tEgeo = 288.8$\,days, $\piEN = -0.033$, and $\piEE = -0.073$. The relative lens-source proper motion vector (in the geocentric frame) has the same direction as the parallax vector. We need to transform it to the barycentric frame. Following \citet{gould2004}:
\begin{equation}
\muhelio = \mugeo + \frac{\ve\pirel}{\au} = \mu \frac{\piEvec}{\piE} +  \frac{\ve\pirel}{\au}.
\label{eq1}
\end{equation}
Here, $\ve = (-2.60,-9.27)\,\mathrm{km}\,\mathrm{s}^{-1}$ is the velocity of the Earth at $\tpar$ (as seen from the solar system barycenter) projected onto the sky at the position of the event, in North and East directions, respectively. The proper motion of the lens (in the barycentric frame) is equal to:
\begin{equation}
\muhelioL = \muhelio + \muhelioS = \mu \frac{\piEvec}{\piE} +  \frac{\ve\pirel}{\au} + \muhelioS,
\label{eq5}
\end{equation}
where $\muhelioS = (-6.50,-4.42)$\,mas\,yr$^{-1}$ is the proper motion of the source, in the North and East directions, respectively (taken from \textit{Gaia} EDR3, \cite{gaia_edr3}, and assuming that the lens is dark, that is, the proper motion of the source is equal to the proper motion of the baseline object). This vector can be transformed to the Galactic coordinate system $(\mu_l,\mu_b)$ using formulae of \citet{poleski2013}.

To evaluate the Galactic prior, we calculate the angular Einstein radius $\thetaE = \tE \mu$, lens mass $M=\thetaE/(\kappa\piE)$ and distance $1/D_l = \thetaE\piE+1/D_s$. We then calculate the tangential velocity of the lens in $l$ and $b$ directions: $V_l = 4.74 \mu_l D_l + V_{\odot}$ and $V_b = 4.74 \mu_b D_l + W_{\odot}$, where $V_{\odot} = 232.2$\,km/s and $W_{\odot}=7.3$\,km/s \citep{schonrich2010} is the velocity of the Sun relative to the Galactic center. To evaluate $f(\muvec)$, we use the following formula:
\begin{equation}
f(\muvec)=f(\mu_l)f(\mu_b) = D^2_l f(V_l) f(V_b),
\end{equation}
where $f(V_l)$ and $f(V_b)$ are normal distributions and the factor $D_l^2$ comes from transforming the proper motion to the tangential velocity.

We use the star density distributions from \citet{batista2011} (which are themselves adapted from \citealt{han_gould2003}). For the disk:
\begin{equation}
\nu_d(R,z) = 1.07\,M_{\odot}\,\mathrm{pc}^{-3}\exp(-R/H)[(1-\beta)\exp(-|z|/h_1)+\beta\exp(-|z|/h_2)],
\label{eq:disk}
\end{equation}
where $H=2.75$\,kpc, $h_1=0.156$\,kpc, $h_2=0.439$\,kpc, and $\beta=0.381$. Here, $R$ is the Galactocentric distance projected onto the Galactic plane and $z$ is the distance from the Galactic plane. For the bulge:
\begin{equation}
\nu_b(r_s) = 1.23\,M_{\odot}\,\mathrm{pc}^{-3}\exp(-0.5r^2_s),
\label{eq:bulge}
\end{equation}
where $r_s^4 = ((x'/x_0)^2+(y'/y_0)^2)^2+(z/z_0)^4$. Here the coordinates ($x',y',z$) have their center at the Galactic center, the longest axis is the $x'$, which is rotated $20^{\circ}$ from the Sun-GC axis toward positive longitude and the $z$ axis points toward the North Galactic pole. The values of the scale lengths are $x_0=1.58$\,kpc, $y_0=0.62$\,kpc, and $z_0=0.43$\,kpc, respectively. The bulge is truncated at $R>2.4$\,kpc from the Galactic center. We also set the limit of the disk range to be $R>1$\,kpc. 

For the velocity distributions, we assume Gaussian distributions as did \citet{han1995} and \citet{batista2011}. For the bulge: $V_l \sim N (0,100)$, $V_b \sim N(0,100)$\,km/s, for disk lenses: $V_l \sim N (220,30)$, $V_b \sim N(0,20)$\,km/s, where $N(X,Y)$ denotes a normal distribution with the mean of $X$ and the standard deviation of $Y$.

\begin{figure}
\includegraphics[width=0.49\textwidth]{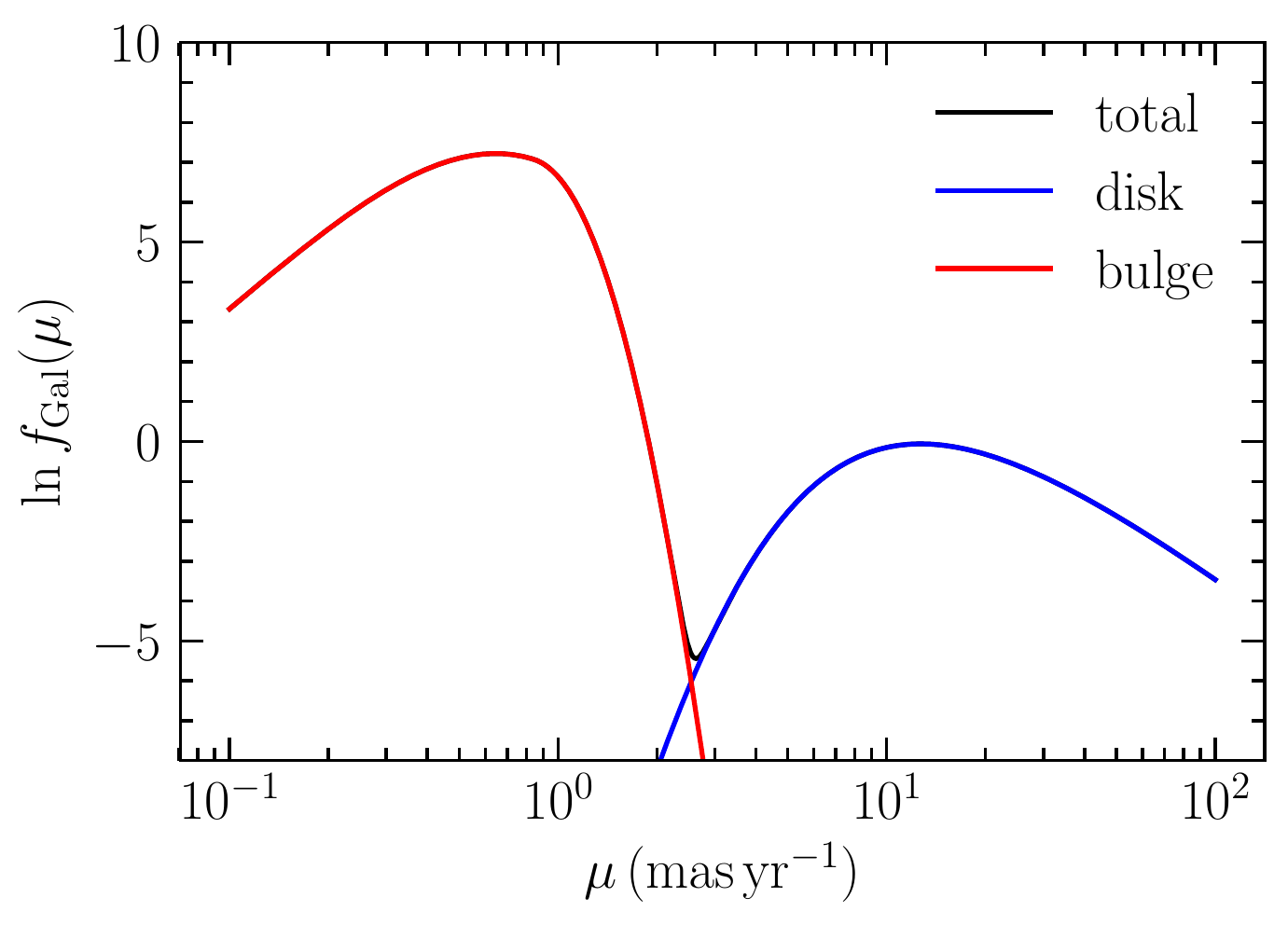}
\includegraphics[width=0.49\textwidth]{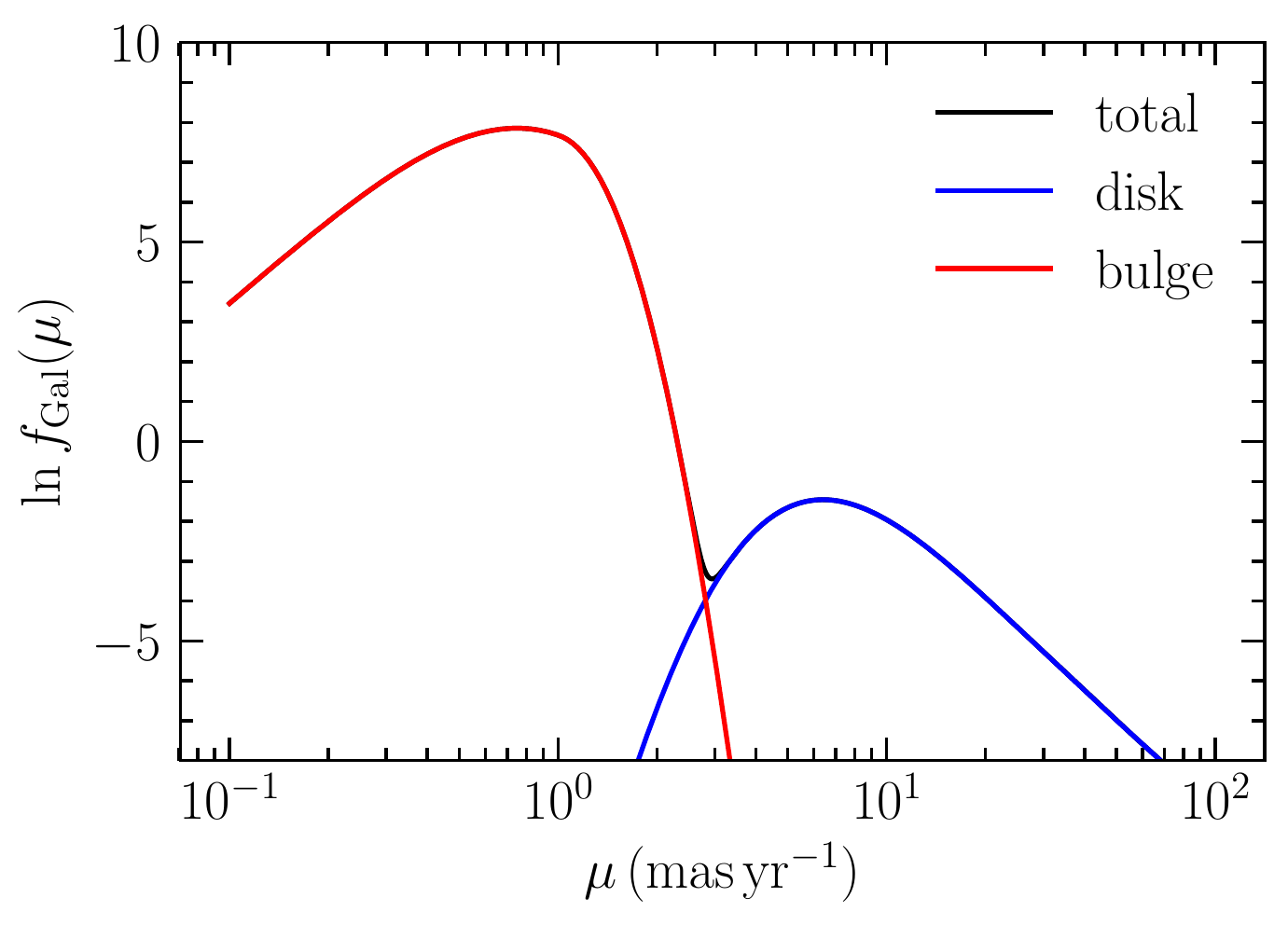}
\caption{Galactic prior on the geocentric relative lens-source proper motion $\mu$ for event OGLE3-ULENS-PAR-02. Other model parameters are taken from \citetalias{wyrzykowski2016} and are held fixed. Left: $u_0<0$ model, right: $u_0>0$ model. In both solutions, Galactic bulge lenses are preferred. However, the prior probability of both degenerate models is similar.}
\label{fig:lnlike}
\end{figure}

Figure~\ref{fig:lnlike} presents the Galactic prior on $\mu$ for the event OGLE3-ULENS-PAR-02 for both possible light curve solutions presented by \citetalias{wyrzykowski2016}. In both cases, Galactic bulge ($\mu<2$\,mas\,yr$^{-1}$) lenses are preferred. 

\subsection{Final models}

In our final models, we allow all eight parameters to vary. The algorithm\footnote{\url{https://github.com/przemekmroz/microlensing_black_holes}} is implemented using the Markov chain Monte Carlo sampler by \citet{foreman2013}. In addition to the ``Galactic prior'' (Equation (\ref{eq:galprior})), we use the two-dimensional normal distribution as a prior on the source proper motion based on \textit{Gaia}~EDR3 measurements\footnote{\textit{Gaia}~EDR3 does not include the proper motion for OGLE3-ULENS-PAR-13 for which we use the \textit{Gaia}~DR2 data.} (taking into account the correlation between $\mu_{s,N}$ and $\mu_{s,E}$). We keep the source distance fixed $D_s=8$\,kpc. (However, in Section~\ref{sec:dist}, we test how source distances calculated based on \textit{Gaia}~EDR3 parallaxes influence the final results.)

In this paper, we analyze 12 light curve models for eight microlensing events that were proposed by \citetalias{wyrzykowski2020b} to contain black hole or ``mass gap'' lenses. In these models almost the entire light comes from the source star. The lens (or unrelated foreground stars) contribute from 0 to 29\% (median 10\%) of the light in the baseline. Low blending is crucial for the correct interpretation of the \textit{Gaia} data. In general, parallaxes and proper motions of microlensing events measured by \textit{Gaia} are weighted average parallaxes and proper motions of the source, lens, and unresolved blends, which are usually difficult to interpret. By requiring that almost the entire light comes from the source star, we ensure that the measured parameters constrain the distance and proper motion of the source.

Results of our calculations, the estimated masses and distances to all lenses, are presented in the fifth and sixth columns of Table~\ref{tab1}. These can be compared to the results of \citetalias{wyrzykowski2020b} (third and fourth columns of Table~\ref{tab1}). We discuss the differences in Section~\ref{sec:discussion}.

\begin{sidewaystable}
\centering
\caption{Posterior estimates on masses and distances of black hole and ``mass-gap'' events from \citetalias{wyrzykowski2020b}.}
\label{tab1}
\begin{tabular}{llrrrrrrrr}
\hline \hline
& & \multicolumn{2}{c}{Wyrzykowski} & \multicolumn{2}{c}{This work} & \multicolumn{2}{c}{This work} & \multicolumn{2}{c}{This work}  \\
& & \multicolumn{2}{c}{\& Mandel (2020)} & \multicolumn{2}{c}{(Fiducial model)} & \multicolumn{2}{c}{(Prior on $D_s$)} & \multicolumn{2}{c}{(Kroupa mass fun.)}  \\
\hline
Lens name & Solution & Mass & Distance & Mass & Distance & Mass & Distance & Mass & Distance \\
OGLE3-ULENS- & ($u_0,\pi_{\rm EN}$) & ($M_{\odot}$) & (kpc) & ($M_{\odot}$) & (kpc) & ($M_{\odot}$) & (kpc) & ($M_{\odot}$) & (kpc) \\
\hline
PAR-02 & $(-\ -)$ &$11.9^{+4.9}_{-5.2}$ & $1.3^{+0.7}_{-0.3}$ &  $0.9^{+0.4}_{-0.3}$  & $5.9^{+0.7}_{-0.6}$ & $1.0^{+0.5}_{-0.4}$  & $6.1^{+0.9}_{-0.8}$ &  $0.8^{+0.4}_{-0.3}$  & $6.1^{+0.6}_{-0.6}$ \\ 
PAR-03 & $(+\ +)$ &$2.4^{+1.9}_{-1.3}$  & $0.6^{+0.5}_{-0.2}$ &  $1.2^{+1.1}_{-0.6}$  & $1.1^{+0.7}_{-0.5}$ & $1.1^{+1.0}_{-0.6}$  & $1.1^{+0.7}_{-0.5}$ &  $1.0^{+0.9}_{-0.5}$  & $1.3^{+0.7}_{-0.6}$  \\ 
PAR-04 & $(+\ +)$ &$2.9^{+1.4}_{-1.3}$  & $0.7^{+0.4}_{-0.2}$ &  $0.6^{+0.7}_{-0.3}$  & $3.1^{+1.4}_{-1.2}$ & $0.7^{+0.9}_{-0.4}$  & $2.5^{+1.3}_{-1.1}$ &  $0.6^{+0.5}_{-0.3}$  & $3.3^{+1.1}_{-1.1}$  \\ 
PAR-04 & $(-\ -)$ &$3.2^{+1.3}_{-1.3}$  & $0.7^{+0.4}_{-0.2}$ &  $0.7^{+1.0}_{-0.4}$  & $2.8^{+1.5}_{-1.3}$ & $1.0^{+1.2}_{-0.6}$  & $2.2^{+1.2}_{-1.0}$ &  $0.7^{+0.7}_{-0.3}$  & $3.0^{+1.2}_{-1.2}$  \\ 
PAR-05 & $(+\ +)$ &$6.7^{+3.2}_{-2.7}$  & $1.6^{+0.7}_{-0.4}$ &  $2.3^{+1.0}_{-0.7}$  & $3.9^{+0.8}_{-0.7}$ & $2.2^{+1.0}_{-0.7}$  & $4.1^{+0.9}_{-0.8}$ &  $2.1^{+0.9}_{-0.7}$  & $4.0^{+0.8}_{-0.7}$  \\ 
PAR-07 & $(+\ -)$ &$4.5^{+1.8}_{-1.8}$  & $1.5^{+0.7}_{-0.3}$ &  $1.2^{+1.4}_{-0.7}$  & $3.7^{+1.4}_{-0.7}$ & $1.1^{+1.3}_{-0.7}$  & $3.9^{+1.9}_{-1.5}$ &  $0.9^{+0.9}_{-0.5}$  & $4.4^{+1.4}_{-1.5}$  \\ 
PAR-13$\dagger$ & $(+\ -)$ &$9.0^{+3.9}_{-3.7}$  & $1.9^{+0.9}_{-0.5}$ &  $5.9^{+2.1}_{-1.7}$  & $3.2^{+0.7}_{-0.6}$ & $6.0^{+2.1}_{-1.8}$  & $3.2^{+0.8}_{-0.6}$ &  $5.4^{+2.1}_{-2.3}$  & $3.4^{+1.0}_{-0.7}$  \\ 
PAR-13$\dagger$ & $(-\ +)$ &$8.0^{+4.1}_{-3.2}$  & $1.8^{+0.8}_{-0.4}$ &  $5.5^{+2.3}_{-1.8}$  & $3.0^{+0.7}_{-0.7}$ & $5.5^{+2.1}_{-1.8}$  & $3.0^{+0.8}_{-0.7}$ &  $5.0^{+2.2}_{-1.9}$  & $3.1^{+0.8}_{-0.7}$  \\ 
PAR-15 & $(+\ -)$ &$3.6^{+1.4}_{-1.6}$  & $1.2^{+0.6}_{-0.3}$ &  $0.8^{+0.6}_{-0.4}$  & $3.7^{+1.0}_{-1.1}$ & $0.8^{+0.6}_{-0.4}$  & $3.6^{+1.2}_{-1.0}$ &  $0.8^{+0.5}_{-0.3}$  & $3.9^{+1.0}_{-1.0}$  \\ 
PAR-15 & $(-\ +)$ &$2.8^{+1.3}_{-1.3}$  & $1.3^{+0.6}_{-0.4}$ &  $0.6^{+0.4}_{-0.3}$  & $4.0^{+1.0}_{-1.0}$ & $0.6^{+0.4}_{-0.3}$  & $3.9^{+1.1}_{-1.0}$ &  $0.6^{+0.4}_{-0.2}$  & $4.1^{+0.9}_{-0.9}$  \\ 
PAR-30$\dagger$ & $(+\ +)$ &$3.0^{+1.3}_{-1.2}$  & $1.3^{+0.6}_{-0.3}$ &  $0.3^{+0.2}_{-0.1}$  & $5.9^{+0.7}_{-0.6}$ & $3.7^{+1.9}_{-1.8}$  & $0.3^{+0.2}_{-0.1}$ &  $0.3^{+0.2}_{-0.1}$  & $5.8^{+0.7}_{-0.6}$  \\ 
PAR-30$\dagger$ & $(-\ -)$ &$2.8^{+1.1}_{-1.1}$  & $1.2^{+0.5}_{-0.3}$ &  $2.3^{+1.7}_{-2.1}$  & $1.1^{+4.5}_{-0.4}$ & $3.7^{+1.9}_{-1.5}$  & $0.3^{+0.2}_{-0.1}$ &  $1.7^{+1.7}_{-1.6}$  & $1.3^{+4.5}_{-0.5}$  \\ 
\hline
\end{tabular}

$\dagger$ \textit{Gaia} astrometric solution is unreliable for this event.
\end{sidewaystable}

\section{Discussion}

\subsection{Prior on the source distance}
\label{sec:dist}

Positions of analyzed source stars in the color--magnitude diagram and their proper motions indicate that they are located in the Galactic bulge and so, in our models, we keep the source distance fixed $D_s=8$\,kpc. \citetalias{wyrzykowski2020b} used distances from \citet{bailerjones2018} as the prior on the source distance. These values are likely underestimated. As explained by \citet{bailerjones2018}, \textit{Gaia} parallaxes are noisy measurements of the inverse of the distance. To correctly interpret parallaxes, one needs to assume a physically motivated prior distribution. \citet{bailerjones2018} adopted the exponentially decreasing space density prior in distance $r$: $f(r) \propto r^2 e^{-r/L}$, where $L$ is a length scale calculated based on the Galactic model. While this prior may be reasonable outside the Galactic plane, it is not appropriate in the present case. Microlensing events analyzed by \citetalias{wyrzykowski2016} occurred on bright giant stars that are visible all the way to the Galactic bulge and behind it. Thus, density priors from Equations~(\ref{eq:disk}) and~(\ref{eq:bulge}) are better suited: $f(r) \propto r^2(\nu_d(r)+\nu_b(r))$. To illustrate the impact of priors on the source distance, let us consider the microlensing event OGLE3-ULENS-PAR-15 (with parallax $0.28 \pm 0.25$\,mas according to \textit{Gaia}~DR2). \citet{bailerjones2018} estimated the source distance of $3.9^{+3.8}_{-1.8}$\,kpc, while our prior indicates that the source is much farther, $7.3^{+2.0}_{-1.7}$\,kpc.

To estimate the impact of uncertain source distances on final lens parameters (masses and distances), we add the source distance as an additional, ninth model parameter and apply the following prior \citep[e.g.,][]{luri2018,bailerjones2018} on it:
\begin{equation}
f(D_s) = D_s^2 (\nu_d(D_s)+\nu_b(D_s)) \times \frac{1}{D_s^2 \sigma_{\varpi}\sqrt{2\pi}}\exp\left(-\frac{(1/D_s-\varpi-\varpi_0)^2}{2\sigma^2_{\varpi}}\right),
\end{equation}
where $\varpi \pm \sigma_{\varpi}$ is the \textit{Gaia} measurement and $\varpi_0$ is the parallax zero point. For the \textit{Gaia}~EDR3 data set, it was shown \citep{lindegren2020} that the parallax zero point varies as a function of magnitude, color, and ecliptic latitude of the object and we calculate the parallax zero points using their recipe. For \textit{Gaia} DR2 (event OGLE3-ULENS-PAR-13), we use $\varpi_0=-0.029$\,mas \citep{lindegren2018}.

The estimated masses and distances to the analyzed lenses, taking into account the \textit{Gaia} parallax and the Galactic prior on density of sources, are presented in the seventh and eight columns of Table~\ref{tab1}. Overall, they are very similar to the parameters measured in our fiducial model with the fixed source distance. Most of \textit{Gaia} parallaxes are consistent with zero and weakly constrain the source distance (so the inferred distances are consistent with the source being in the Galactic bulge, where the prior density of stars is the largest). The only exception is the microlensing event OGLE3-ULENS-PAR-30 which seems to be caused by a lens that is more massive and located closer than in the fiducial model. Its \textit{Gaia} EDR3 parallax ($\varpi = 2.98 \pm 0.58$\,mas) suggests that the source is nearby, however, this astrometric solution is likely spurious (its renormalized unit weight error\footnote{The RUWE is expected to be around 1 for sources with astrometric solution that describes data well. RUWE values significantly greater than 1 could indicate an incorrect astrometric solution \citep{lindegren2021}.}, RUWE, is 2.75). We conclude that \textit{Gaia}~EDR3 parallaxes of analyzed microlensing events do not strongly constrain source distances or may even lead to incorrect results.

\subsection{Prior on the lens mass function}

In our fiducial model, following \citetalias{wyrzykowski2016}, we assume a power-law lens mass function $g(M)\propto M^{-1.75}$. To check how the choice of the mass function affects the inferred lens mass and distance, we re-run the models using the \citet{kroupa2001} mass function as our prior:
\begin{equation}
g(M) = \begin{cases}
a_1 M^{-0.3} & \text{for $0.01 < M < 0.08\,M_{\odot}$} \\
a_2 M^{-1.3} & \text{for $0.08 < M < 0.5\,M_{\odot}$} \\
a_3 M^{-2.3} & \text{for $0.5 < M < 150\,M_{\odot}$} \\
\end{cases},
\end{equation}
where $a_1=1.9870$, $a_2=0.1590$, and $a_3=0.0795$. This mass function is steeper than that used by \citetalias{wyrzykowski2016} for objects more massive than $0.5\,M_{\odot}$. Results of our calculations are reported in the ninth and tenth columns of Table~\ref{tab1}. Unsurprisingly, the choice of the mass function has little impact on final results. The only exception is again the microlensing event OGLE3-ULENS-PAR-30 for which disk and bulge lenses have similar prior probabilities so modifying the mass function slightly changes the relative probability of disk and bulge solutions. The inferred masses and distances are, however, statistically consistent. (Note that the \textit{Gaia} astrometric solution is unreliable in this case.)

The mass function of isolated stellar remnants is poorly known and actually the goal of the analyses of \citetalias{wyrzykowski2016} and \citetalias{wyrzykowski2020b} was to measure it. We demonstrate that the inferred lens masses are weakly sensitive to the assumed mass function. In a companion paper (Mróz et al. 2021, in prep.), we present new constraints on the remnant mass function derived from a sample of over 4000 microlensing events detected by the OGLE-III survey.

\subsection{Natal kicks}

\begin{figure}
\centering
\includegraphics[width=0.49\textwidth]{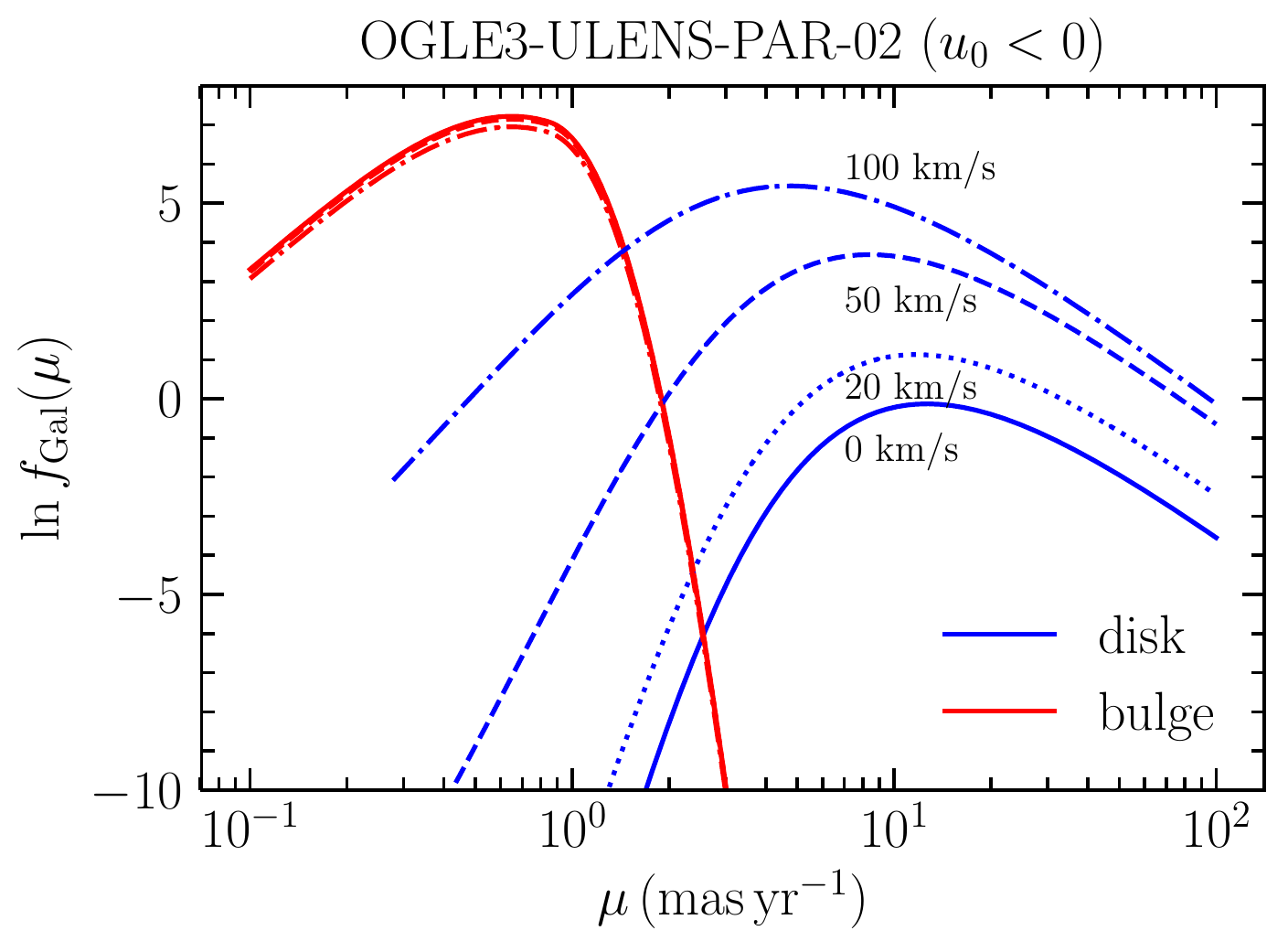}
\includegraphics[width=0.49\textwidth]{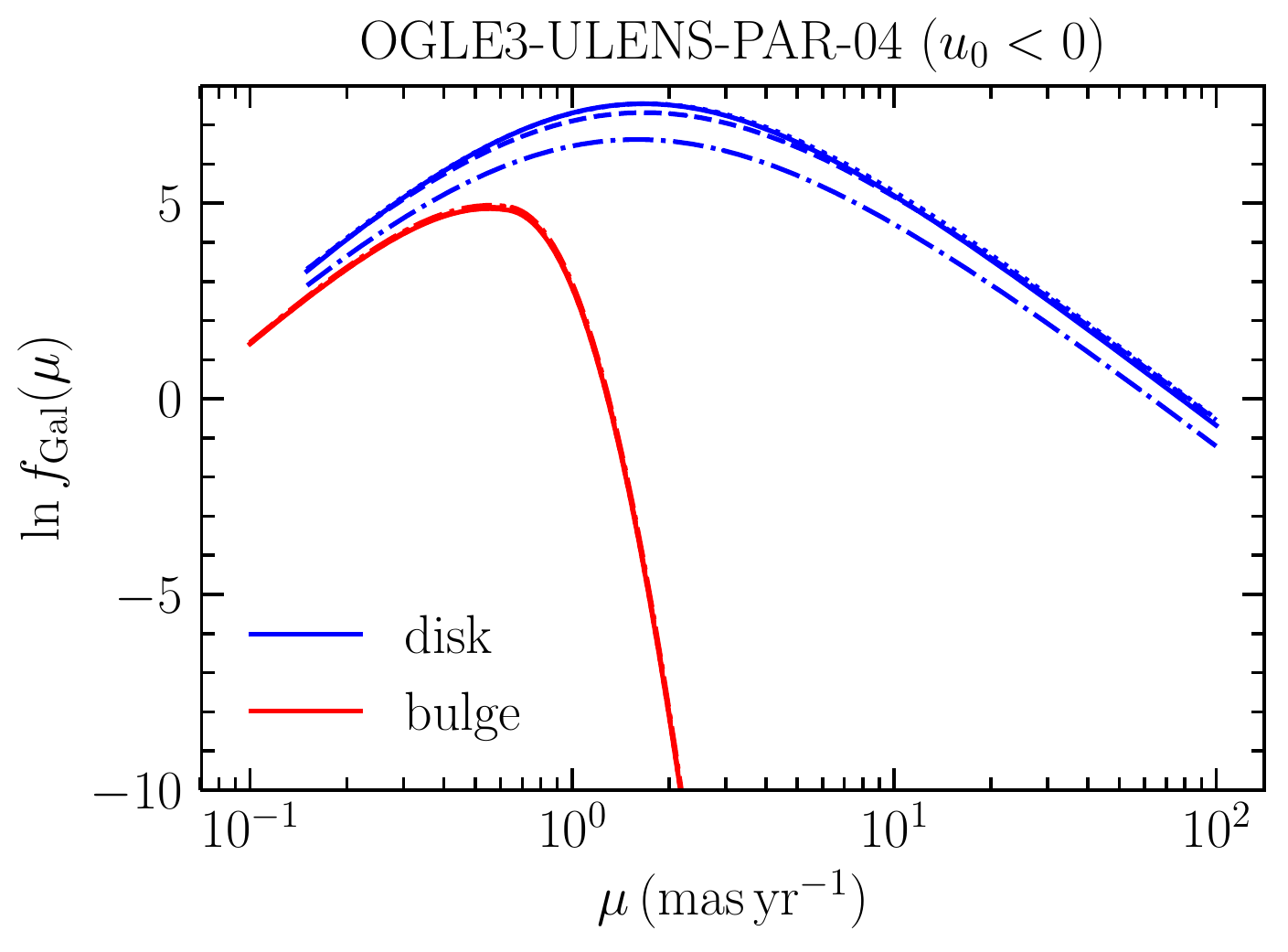}
\caption{Galactic prior on the geocentric relative lens-source proper motion $\mu$ for events OGLE3-ULENS-PAR-02 ($u_0<0$ model) and OGLE3-ULENS-PAR-04 ($u_0<0$ model), assuming different natal kick velocities for the entire lens population. Solid, dotted, dashed, and dashed-dotted lines correspond to  $v_{\rm kick} = 0,\ 20,\ 50,\ \mathrm{and}\ 100\ \mathrm{km\,s}^{-1}$, respectively.}
\label{fig:kicks}
\end{figure}

All lens masses are calculated under the assumption that the distribution of velocities of lenses is consistent with that of stars in the Milky Way. However, it is known that neutron stars may receive large natal kicks \citep{hobbs2005}. The formation of black holes is less understood with some studies arguing that (at least some) black holes receive small natal kicks \citep[e.g.][]{miller2009,wong2012}, while other studies favor large ($\sim 100$\,km\,s$^{-1}$) kick velocities \citep[e.g.,][]{willems2005,repetto2012,repetto2015,atri2019}.

The direction of the kick velocity vector with respect to the Galactic coordinates is random; for any individual event, the relative lens-source proper motion may be larger or smaller than the typical proper motion in that direction of the sky. However, the net effect is that the population of stellar remnants may have a larger velocity dispersion compared to that of other stars. Assuming that remnants receive a kick velocity of $v_{\rm kick}$, the velocity dispersion of remnants (in $l$ and $b$ directions) is $\sigma^2= \sigma_0^2 + v^2_{\rm kick}/3$, where $\sigma_0^2$ is the velocity dispersion in a given direction. Thus, natal kicks may affect the derived masses only if $v_{\rm kick} \gtrsim \sigma_0$.

Figure~\ref{fig:kicks} shows how our Galactic prior on the lens-source proper motion depends on $v_{\rm kick}$ assuming that the \textit{entire} population of lenses receives the same kick velocity. Solid, dotted, dashed, and dashed-dotted lines correspond to  $v_{\rm kick} = 0,\ 20,\ 50,\ \mathrm{and}\ 100\ \mathrm{km\,s}^{-1}$, respectively. Because $\sigma_0 \approx 100$\,km\,s$^{-1}$ in the bulge and $\sigma_0 < v_{\rm kick}$, the likelihood that the lens is the bulge, in practice, does not change. However, the velocity dispersion in the Galactic disk is much lower ($\sigma_0=20-30$\,km\,s$^{-1}$), so natal kicks may affect both the total probability that the lens is located in the disk as well as the most likely value of the lens-source proper motion (and, by the same token, the most likely lens mass). The effect of natal kicks depends on the geometry of a given event (especially the direction of the parallax vector) and in many cases kicks do not significantly influence the most likely value of $\mu$ (as presented in the right panel of Figure~\ref{fig:kicks}).

\section{Summary}
\label{sec:discussion}

The inferred masses of candidate dark lenses (Table~\ref{tab1}) are $\approx 2-12$ times smaller than those found by \citetalias{wyrzykowski2020b}. As explained in Section \ref{sec:intro}, this discrepancy is caused by a combination of factors. First, \citetalias{wyrzykowski2016} and \citetalias{wyrzykowski2020b} assumed that lenses are located in the Galactic disk and neglected bulge lenses. They argued that the analyzed events exhibit a clear microlensing signal and a large parallax indicates a nearby lens, whereas bulge lenses would have small and hardly detectable parallaxes. This is certainly true for the majority of events detected by \citetalias{wyrzykowski2016} but not all. As presented in Figure \ref{fig:lnlike}, OGLE3-ULENS-PAR-02 is likely located in the (near) bulge; OGLE3-ULENS-PAR-30 may be located either in the disk or in the bulge with similar prior probabilities.

Another crucial assumption made by \citetalias{wyrzykowski2020b} was using underestimated source distances taken from the \citet{bailerjones2018} catalog. The median source distances of eight events reanalyzed in this paper were assumed to be in the range from 2.6 to 5.9\,kpc with the median of 4.0\,kpc whereas the actual \textit{Gaia} measurements are consistent with the source located in the Galactic bulge at 8\,kpc. The Galactic prior used by \citet{bailerjones2018} is not appropriate for bright giant sources located toward the Galactic bulge and the use of incorrect prior leads to underestimated distances.

To measure the posterior distribution of the lens mass and distance, \citetalias{wyrzykowski2016} and \citetalias{wyrzykowski2020b} assigned every link of the MCMC chain a random value of the lens-source proper motion $\mu$. They sampled both components of $\muvec$ from uniform distributions so the distribution of $\mu=|\muvec|$ was biased toward higher proper motions (and so larger lens masses), which also contributed to the overestimation of masses of ``mass gap'' and black hole lenses.

Our re-analysis of \citetalias{wyrzykowski2020b} data indicates that all their ``mass gap'' and black hole lenses likely have much lower masses that are consistent with white dwarf and neutron star lenses (Table~\ref{tab1}). The only exception appears to be OGLE3-ULENS-PAR-13 with the inferred mass of $5.9^{+2.1}_{-1.7}\,M_{\odot}$. However, this measurement is based on an unreliable \textit{Gaia} proper motion measurements of the source star (the proper motion of that star is not included in \textit{Gaia}~EDR3, whereas the \textit{Gaia}~DR2 astrometric solution has a relatively high RUWE of 1.67). An alternative proper motion measurement of that star is available from OGLE (Mróz et al. 2021, in prep.) and yields a lower lens mass of $2.9^{+2.0}_{-1.8}\,M_{\odot}$.

All inferred masses have only a statistical meaning -- these masses correspond to the most likely value of the relative lens-source proper motion in the adopted Galactic model. We cannot exclude that some of the analyzed lenses have unusual kinematics and so may be indeed black holes. However, the most likely explanation is that these objects are main sequence stars, white dwarfs, or neutron stars.

\section*{Acknowledgements}

We thank Ilya Mandel, Radek Poleski, Jan Skowron, and Andrzej Udalski for discussions and their comments on the manuscript. P.M.~acknowledges the support from Heising-Simons Foundation Grant \#2018-1036 awarded to J. Fuller. {\L}.W.~acknowledges support from the Polish National Science Center grants: Harmonia No.~2018/30/M/ST9/00311 and Daina No.~2017/27/L/ST9/03221. This work has made use of data from the European Space Agency (ESA) mission {\it Gaia} (\url{https://www.cosmos.esa.int/gaia}), processed by the {\it Gaia} Data Processing and Analysis Consortium (DPAC, \url{https://www.cosmos.esa.int/web/gaia/dpac/consortium}). Funding for the DPAC has been provided by national institutions, in particular the institutions participating in the {\it Gaia} Multilateral Agreement.

\bibliographystyle{aasjournal}
\bibliography{pap}

\end{document}